\documentclass[a4paper,11pt]{article}
\usepackage{pos}
\usepackage{booktabs}
\usepackage{xcolor}
\usepackage{graphicx}
\usepackage{todonotes}
\title{How, where and when do cosmic rays reach ultrahigh energies?}

\author*[a]{James H. Matthews}
\author[b]{Andrew M. Taylor}

\affiliation[a]{Department of Physics, Astrophysics, University of Oxford, Denys Wilkinson Building, Keble Road, Oxford, OX1 3RH, UK}
\affiliation[b]{Deutsches Elektronen-Synchrotron, Platanenallee 6, Zeuthen, Germany}

\emailAdd{james.matthews@physics.ox.ac.uk}

\abstract{Understanding the origins of ultrahigh energy cosmic rays (UHECRs) -- which reach energies in excess of $10^{20}$ eV -- stretches particle acceleration physics to its very limits. In this review, we discuss {\em how} such energies can be reached, using general arguments that can often be derived on the back of an envelope. We explore possible particle acceleration mechanisms, with special attention paid to shock acceleration. Informed by the arguments derived, we discuss {\em where} UHECRs might come from and which classes of powerful astrophysical objects could be UHECR sources; generally, we favour radio galaxies, GRB afterglows and other sources which are not too compact and dissipate prodigious amounts of energy on large scales, allowing them to generate large products $\beta B R$ without the CRs undergoing restrictive losses. Finally, we discuss {\em when} UHECRs are accelerated by highlighting the importance of source variability, and explore the intriguing possibility that the UHECR arrival directions are partly a result of ``echoes'' from magnetic structures in the local Universe.}

\FullConference{%
  27th European Cosmic Ray Symposium - ECRS \\
  25-29 July 2022 \\
  Nijmegen, the Netherlands 
}

\newcommand\vel{v}

\begin{document}
\def\aj{Astron. J.}
\def\actaa{Acta Astron.}
\def\araa{Annu. Rev. Astron. Astrophys.}
\def\apj{Astrophys. J.}
\def\apjl{Astrophys. J. Lett.}
\def\apjs{Astrophys. J. Suppl.}
\def\ao{Appl. Opt.}
\def\apss{Astrophys. Space Sci.}
\def\aap{Astron. Astrophys.}
\def\aapr{Astron. Astrophys. Rev.}
\def\aaps{Astron. Astrophys. Suppl.}
\def\azh{Astronomicheskii Zhurnal}
\def\baas{Bulletin of the AAS}
\def\bac{Bulletin of the Astronomical Institutes of Czechoslovakia}
\def\caa{Chinese Astronomy and Astrophysics}
\def\cjaa{Chinese Journal of Astronomy and Astrophysics}
\def\icarus{Icarus}
\def\jcap{JCAP}
\def\jrasc{Journal of the RAS of Canada}
\def\memras{Memoirs of the RAS}
\def\mnras{Mon. Not. Roy. Astron. Soc.}
\def\na{New Astronomy}
\def\nar{New Astronomy Review}
\def\pra{Phys. Rev. A: General Physics}
\def\prb{Phys. Rev. B: Solid State}
\def\prc{Phys. Rev. C}
\def\prd{Phys. Rev. D}
\def\pre{Phys. Rev. E}
\def\prl{Phys. Rev. Lett.}
\def\pasa{Publications of the Astron. Soc. of Australia}
\def\pasp{Publications of the ASP}
\def\pasj{Publications of the ASJ}
\def\rmxaa{Revista Mexicana de Astronomia y Astrofisica}
\def\qjras{Quarterly Journal of the RAS}
\def\skytel{Sky and Telescope}
\def\solphys{Solar Physics}
\def\sovast{Soviet Astronomy}
\def\ssr{Space Science Reviews}
\def\zap{Zeitschrift für Astrophysik}
\def\nat{Nature}
\def\iaucirc{IAU Cirulars}
\def\aplett{Astrophys. Lett.}
\def\apspr{Astrophysics Space Physics Research}
\def\bain{Bulletin Astronomical Institute of the Netherlands}
\def\fcp{Fundamental Cosmic Physics}
\def\gca{Geochimica Cosmochimica Acta}
\def\grl{Geophysics Research Letters}
\def\jcp{Journal of Chemical Physics}
\def\jgr{Journal of Geophysics Research}
\def\jqsrt{Journal of Quantitiative Spectroscopy and Radiative Transfer}
\def\memsai{Mem. Societa Astronomica Italiana}
\def\nphysa{Nuclear Physics A}
\def\physrep{Physics Reports}
\def\physscr{Physica Scripta}
\def\planss{Planetary Space Science}
\def\procspie{Proceedings of the SPIE}

\let\astap=\aap
\let\apjlett=\apjl
\let\apjsupp=\apjs
\let\applopt=\ao

\def\ndash{--}
\def\mdash{---}
\maketitle

\section{Introduction}
The origin of ultrahigh energy cosmic rays (UHECRs) has remained an open question ever since their discovery by Linsley \cite{linsley_evidence_1963}. Together with Scarsi, Linsley measured the energy spectrum above $1$~EeV and even performed an analysis of arrival directions with a sample of 97 events. Since these pioneering results, decades of experimental and theoretical work have been dedicated to understanding the phenomenology and physics of UHECRs (see historical reviews by Watson \cite{watson_highest-energy_2019,watson_origins_2019}) -- {\em the highest energy particles in nature}. Despite Herculean efforts, the sources of UHECRs are not yet known, nor is the physics of their acceleration understood. 

The current state of the art UHECR observatories are the Pierre Auger Observatory (PAO), in 	Malargüe, Argentina (detection area $\approx3000~{\rm km}^2$), and the Telescope Array (TA) in  Millard County, Utah, USA (detection area $\approx700~{\rm km}^2$). Both observatories have been critical for measuring the spectrum, composition and anisotropy of UHECRs over the past decade. The CR spectrum is characterised by a smooth power-law over $11$ decades in energy, with a series of inflection points; in the UHE regime the most relevant features are the ankle, a hardening at $\approx 4~{\rm EeV}$, and a cutoff or flux suppression at $\approx 40~{\rm EeV}$ \cite[e.g.][see Fig.~\ref{fig:loss_lengths}]{pierre_auger_collaboration_combined_2017}. In terms of composition, combined fits of the spectrum and the distribution or moments of the depths of the air shower maxima, $X_{\rm max}$, suggest a composition that gets heavier with energy above the ankle \cite[][see also section~\ref{sec:composition}]{pierre_auger_collaboration_combined_2017,bellido_depth_2021,watson_further_2022}. Finally, the question of UHECR anisotropy has seen particularly exciting recent progress, with PAO reporting a dipole anisotropy at $5.2\sigma$ significance \cite{pierre_auger_collaboration_observation_2017}, together with less significant indications of anisotropies on smaller scales from both PAO \cite{pierre_auger_collaboration_indication_2018,pierre_auger_collaboration_arrival_2022} and TA \cite{abbasi_indications_2014,telescope_array_collaboration_indications_2021}. However, working backwards from arrival directions to uncover the sources of UHECRs remains challenging given the limited statistics at such high energies, uncertain detailed composition and, in particular, the obfuscating effect of (poorly constrained) intergalactic and Galactic magnetic fields. 

On the theoretical and modelling side, there have also been many recent advances (see, e.g., chapters 5 \& 6 of the EuCAPT white paper \cite{alves_batista_eucapt_2021}). Building on the foundational theory of shock acceleration \cite{bell_acceleration_1978,blandford_particle_1978,axford_acceleration_1977,krymskii_regular_1977}, particle-in-cell (PIC) simulations have provided unprecedented insights into the nonlinear plasma physics at work during shock acceleration \cite[e.g.][]{spitkovsky_particle_2008,caprioli_simulations_2014} and magnetic reconnection \cite[e.g.][]{drake_electron_2006,sironi_relativistic_2014}. Cosmic-ray propagation codes, such as CR-Propa \cite{alves_batista_crpropa_2016}, now provide flexible frameworks for treating the propagation of UHECRs from source to detector. This mature and well-tested suite of computational tools are essential for understanding the theoretical cosmic-ray landscape, but one particular feature of UHECR acceleration is the vast range of scales at work; a mildly relativistic proton with $\gamma\sim 10$ must increase it's energy (and thus Larmor radius for a constant magnetic field strength) by a factor of $10^8$ to reach the UHE ($\gtrsim 10^{18}~{\rm eV}$) regime. Such a dynamic range is out of reach of even the most ambitious simulator. 

It is not yet possible to make unambiguous inferences about UHECR sources from data or theory alone. As a result, we must consider the whole picture, taking into account plasma physics, astrophysics and multimessenger astronomy, when interpreting the experimental data. The need for such a holistic view makes studying UHECRs challenging, but also particularly rich and rewarding (in our opinion!). We will try to convey some of that excitement in this review contribution to the proceedings of the European Cosmic Ray Symposium (ECRS) 2022. Our review is structured as follows, mirroring the corresponding ECRS talk. We start (section~\ref{sec:fundamentals}) by going over some UHECR fundamentals, to establish the basic assumptions we will make. We then discuss each of the interrogatives in our title: we explore {\em how} UHECRs might be able to reach such extreme energies (section~\ref{sec:how}), {\em where} they might be coming from (section~\ref{sec:where}), and {\em when} they might have been accelerated, with a biased focus on a specific `echoes' model for their origin (section~\ref{sec:when}). Finally, in section~\ref{sec:conclusions}, we conclude and comment on the future outlook. We generally adopt CGS units and Gaussian units for electromagnetism, but we often give energies in eV or EeV and rigidities in V or EV. We use the symbols ${\cal E}$ for electric field, $E$ for energy, $\boldsymbol{\vel}$ for velocity, and define $\beta\equiv \vel/c$.

\section{UHECR Fundamentals}
\label{sec:fundamentals}
\begin{figure}
\centering
\includegraphics[width=\linewidth]{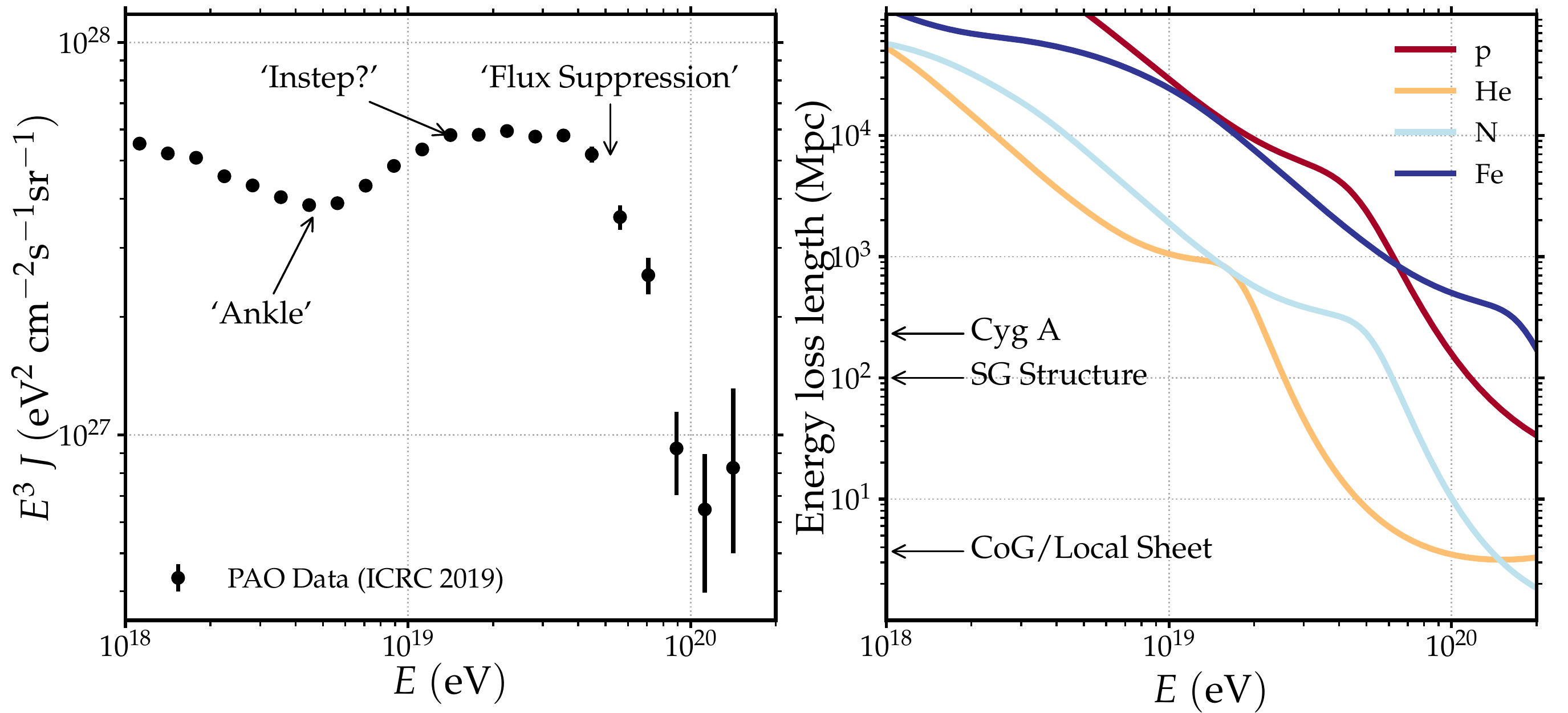}
\caption{
{\sl Left:} The UHECR energy spectrum measured by PAO as presented at ICRC 2019 \cite{pierre_auger_collaboration_pierre_2019}. The spectrum is shown in $E^3~J$ units where $J$ is the differential flux spectrum in units of particles per unit area per unit energy per unit solid angle. As plotted, a $dN/dE \propto E^{-3}$ spectrum shows up as a horizontal line. 
{\sl Right:} Energy loss lengths as a function of CR energy calculated for the same four species shown in Fig.~\ref{fig:xmax}. The energy loss length is defined in the text. The loss lengths are calculated by considering photopion, photodisintegration and pair production interactions with the CMB and EBL, using the EBL model of \cite{franceschini_extragalactic_2008}. 
}
\label{fig:loss_lengths}
\end{figure}

We define ultrahigh energy cosmic rays (UHECRs) as charged particles (protons or nuclei) reaching an energy in excess of $10^{18}$~eV, although a successful UHECR source must be able to accelerate particles right up to $\sim 100~{\rm EeV}$ in order to explain the full energy range of UHECR data. The spectrum of UHECRs arriving at Earth as measured by PAO is shown in the left hand panel of Fig.~\ref{fig:loss_lengths}, with the main features labelled.

The Larmor radius (or gyroradius) of such an ultra-relativistic particle with energy $E=pc$ (where $p$ is momentum) is given by 
$r_g = E/(ZeB)$. This is the radius of gyration when undergoing circular rotation in a uniform magnetic field. Writing down the Larmor radius already gives us useful insights. Given in scaling relation form for characteristic UHECR energies, the Larmor radius is 
\begin{equation}
    r_g = 10.8~{\rm kpc}~\left(\frac{E}{10~{\rm EeV}}\right)~\left(\frac{B}{\mu {\rm G}} \right)^{-1}~Z^{-1}.
    \label{eq:larmor}
\end{equation}
This gives a (very minimal) condition for UHECR sources -- a source must be able to confine a particle before it can accelerate it. However, for acceleration to take place, it is the electric field that really matters, and it is useful to think of the maximum rigidity, rather than maximum energy, associated with astrophysical acceleration sites. We define rigidity, which we quote in Volts (V), as 
\begin{equation}
    {\cal R} = \frac{E}{Ze}.
\end{equation}
The maximum rigidity, ${\cal R}_{\rm max}$, is an important quantity because it relates directly to particle acceleration in electric or magnetic fields. The energy gained by moving a particle a distance $R$ in an electric field of strength ${\cal E}$ is $Ze{\cal E}R$, so the maximum rigidity should be an intrinsic quantity of the accelerator; it depends only on the size of the region, and the electric field available.

\subsection{UHECR losses, propagation and `horizons'}
\label{sec:losses}
UHECRs are attenuated or degraded from interaction with the cosmic microwave background (CMB) and extragalactic background light (EBL). Protons can undergo resonant photopion conversion, which in this context is known as the Greisen-Zat'sepin-Kuzmin (GZK) effect \cite{greisen_end_1966,zatsepin_upper_1966}, while heavier nuclei undergo photodisintegration \cite{stecker_photodisintegration_1999} and all nuclei are subject to Bethe-Heitler pair production. The photopion, pair production and photodisintegration processes impose composition-dependent energy loss lengths or mean free paths, making it difficult for UHECRs to reach us from very distant sources. These length-scales are often referred to as limits or horizons, but as with any opacity source there is a chance, however slim, that an UHECR can travel a considerable distance beyond this length scale. The energy loss lengths for protons and a few different ion species (He, N, Fe) are shown in Fig.~\ref{fig:loss_lengths} for the EBL model of \cite{franceschini_extragalactic_2008}. The energy loss length here is defined as the average distance necessary for an UHECR to propagate in order for its energy to decrease to $e^{-1}$ of its original value. The separate bumps in the curves are attributed to different processes and radiation fields, with the CMB photopion and photodisintegration processes dominating at the highest energies. We have also marked on the figure the distance to Cygnus A of $\sim240~{\rm Mpc}$ \cite{carilli_cygnus_1996}, the typical distance to objects in the `Council of Giants' (CoG) or `Local Sheet' \cite{mccall_council_2014}, and the characteristic scale length of $\sim 100~{\rm Mpc}$ associated with the supergalactic plane \cite{bohringer_cosmic_2016}; these distances are all relevant to discussions here and in section~\ref{sec:where}.  
Fig.~\ref{fig:loss_lengths} highlights how difficult it is to accurately characterise the UHECR source population from the spectrum alone given that its form depends on the source spectrum, spatial distribution or redshift evolution, and composition. Nevertheless, broadly speaking, the various CMB and EBL interactions impose a characteristic length scale $\sim 100$~Mpc within which the dominant UHECR sources are most likely to lie.    

Combined knowledge of source timescales, UHECR propagation and anisotropy can impose additional constraints on UHECR source distances. For example, Eichmann and collaborators have explored a model where Cygnus A was the dominant source in the sky up to tens of EeV \cite{eichmann_ultra-high-energy_2018}. Cygnus A is compelling as an UHECR source because it is unusually powerful for a radio galaxy, and although its distance of $\sim 240$~Mpc  might appear restrictive, at $10^{19}~{\rm eV}$ energy loss lengths are quite large, and Cygnus A's potentially vast UHECR luminosity could still produce the magnitude of the observed UHECR flux. However, in a follow-up paper \cite{eichmann_ultra-high-energy_2019}, Eichmann showed that Cygnus A cannot account for an isotropic CR component at these energies, because the CRs would not have had time to isotropise in the extragalactic magnetic field in the time the source has been active; one should instead see an anisotropic signal pointing towards Cygnus A. This difficulty could in principle be alleviated if the Galactic halo magnetic field can isotropise the signal on shorter timescales, but the general principle of `diffusive' horizons for UHECR production is nevertheless important (see also Refs. \cite{alves_batista_magnetic_2014,globus_cosmic_2019}).  

Following the cumulative composition and spectral measurements made by the PAO over the last 15 years, a growing body of evidence has amounted suggesting that UHECR at the highest energies must have a rather local origin \cite{2015PhRvD..92f3011T,lang_revisiting_2020}. This finding is particularly interesting along with other suggestions that not many sources should be contributing to the UHECR spectrum in this high energy range \cite{Ehlert:2022jmy}. Collectively this suggests that a local UHECR source may dominate the contribution to the UHECR spectrum at the highest energies. This finding may also be consistent with the UHECR dipole strength recently detected by the PAO \cite{PierreAuger:2017pzq}, with the main contribution to the dipole being driven by the presence of this local source \cite{Lang:2020qhh}.

\begin{figure}
\centering
\includegraphics[width=\linewidth]{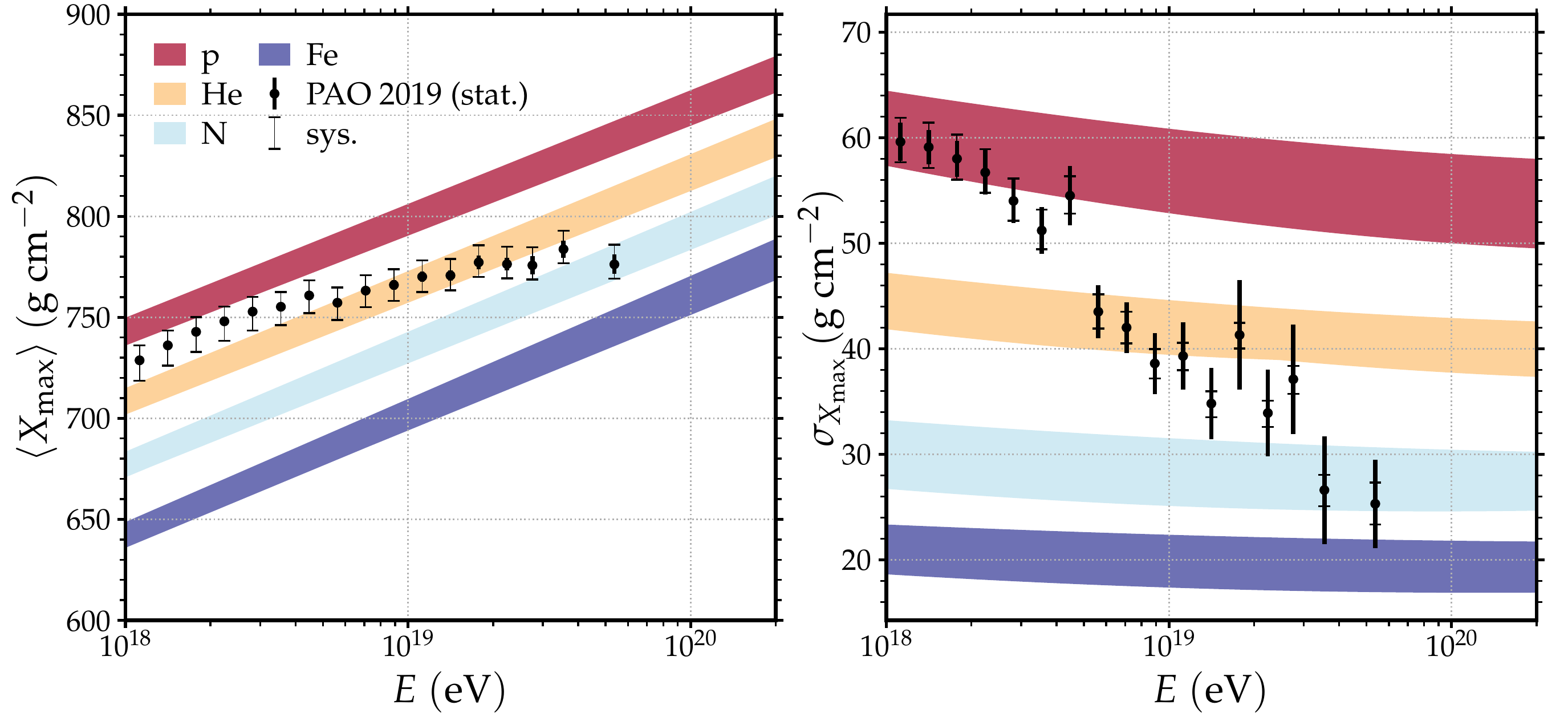}
\caption{
Composition diagnostics from extensive air showers detected by the PAO, showing how the composition gets heavier at higher energies. Data taken from the PAO contribution to ICRC 2019 \cite{pierre_auger_collaboration_pierre_2019}. {\sl Left:} The $\langle X_{\rm max} \rangle$ distribution from PAO data, defined as the mean depth of the air shower maximum, as a function of energy.  {\sl Right:} $\sigma_{X_{\rm max}}$, the standard deviation of $X_{\rm max}$ as a function of energy. In both cases the coloured bands show the predictions for four pure composition scenarios, with the range shown corresponding to that spanned by three hadronic interaction models (QGSJet II-04, EPOS-LHC and Sibyll 2.1), as calculated using the parameterisation from Ref. \cite{pierre_auger_collaboration_interpretation_2013}, with adjusted parameters from S. Petrera (priv. comm.). 
}
\label{fig:xmax}
\end{figure}

\subsection{Composition and Maximum Rigidity}
\label{sec:composition}
Since the maximum rigidity, rather than maximum energy, is an intrinsic property of a cosmic acceleration site, it follows that the charge on the nuclei (or the atomic composition of UHECRs) is important for establishing possible UHECR sources. At CR energies below the knee, experiments such as the {\sl Alpha Magnetic Spectrometer} \cite[AMS;][]{aguilar_alpha_2021} and {\sl Cosmic Ray Energetics and Mass} experiment \cite[CREAM;][]{seo_cosmic-ray_2004} can provide direct measurements of CR charge and therefore decompose the CR spectrum into different species. However, at ultrahigh energies, the CRs are detected through extensive air showers, and the main diagnostic of composition is a more indirect measure: the distribution of the depths of air shower maxima, $X_{\rm max}$.  We show the first two moments of the $X_{\rm max}$ distribution in Fig.~\ref{fig:xmax} from the data released as part of the PAO contribution to ICRC 2019. The data are compared to theoretical distributions for $X_{\rm max}$ for three different hadronic interaction models. The general trend is a gradual change from nearly pure protons around $\sim 3~{\rm EeV}$ to a heavier composition at higher energies. Such a trend might suggest maximum rigidities of ${\cal R}_{\rm max} \sim 3-10~{\rm EV}$, which is broadly consistent with other studies: the combined fit of the spectrum and composition PAO data finds a cutoff rigidity of ${\cal R}_{\rm max} = 4.79~{\rm EV}$ \cite{pierre_auger_collaboration_combined_2017}, and Ref.~\cite{taylor_indications_2015} find maximum energies for Fe nuclei at source of $\approx 300~{\rm EeV}$ suggesting ${\cal R}_{\rm max} \approx 11~{\rm EV}$. There is some wiggle room in this quantity but it cannot be too much lower than 10 EV in order to explain the observed $> 100~{\rm EeV}$ UHECRs, so we will adopt $10~{\rm EV} (10^{19}~{\rm V})$ as our `target' rigidity when discussing UHECR sources. 

\section{Acceleration of UHECRs ({\em How?})}
\label{sec:how}

\subsection{The Hillas energy and power requirement}
The maximum characteristic energy associated with a particle acceleration process is the Hillas energy \cite{hillas_origin_1984}, given by
\begin{equation}
    E_H = 9.25~{\rm EeV}~\left(\frac{B}{10~{\rm \mu G}}\right) \left(\frac{R}{{\rm kpc}}\right)~Z\beta,
    \label{eq:hillas}
\end{equation}
where $Z$ is the dimensionless charge on the particle, $\beta=\vel/c$ is the velocity of the accelerator in units of $c$, $B$ is the magnetic field and $R$ is the characteristic size. The above equation can also be equivalently written as a Hillas rigidity in the form ${\cal R}_H = \beta B R$, which is the basic {\sl figure of merit} for an UHECR accelerator. The Hillas condition is {\em not} the same as having a Larmor radius equal to the size of the acceleration region; there is an additional factor such that the acceleration region must be larger than the Larmor radius of the highest energy particles by a factor of $1/\beta$. The Hillas energy can be arrived at in various ways, but is perhaps best understood in terms of a particle travelling a distance $R$ in an optimally arranged $-(\boldsymbol{\vel}/c) \times \boldsymbol{B}$ electric field. It is is only a characteristic maximum energy, and as we shall see in the next section is a {\em necessary, but not sufficient} criterion that is only reached under certain conditions. One can construct scenarios in which the Hillas energy is exceeded; for example, if a CR can be confined to a perpendicular shock for a very long time then in principle the CR can cross the shock on many occasions without escaping. However, in practice this is likely to require specialised shock or magnetic field geometries to avoid drifts, diffusion or advection removing the particle from the acceleration site. We will therefore proceed under the expectation that the Hillas energy really is a maximum or cutoff energy. 

The Hillas energy can be used to derive a minimum magnetic or kinetic power that a source must possess. A similar requirement was, to our knowledge, first discussed by Lovelace \cite{lovelace_dynamo_1976}, but also forms the basis for Hillas' figure 6 in the 1984 paper \cite{hillas_origin_1984}. The power requirement is therefore sometimes referred to as a `Hillas-Lovelace limit' (although see also Refs. \cite{blandford_acceleration_2000,waxman_high-energy_2001,waxman_high-energy_2004,matthews_particle_2020,rieger_active_2022}). The basic idea is that a source must be able to supply enough magnetic energy per unit time that a given product $\beta B R$ can be maintained in the acceleration site. In the non-relativistic case, a limit on the kinetic power $Q_k$ can be derived, 
\begin{equation}
    Q_k \gtrsim 10^{44}~{\rm erg~s}^{-1}~\beta^{-1}~\left(\frac{E/Z}{10^{19}{\rm eV}} \right)^2~\epsilon_b~\eta^2,
    \label{eq:power}
\end{equation}
where $\epsilon_b$ is the ratio of magnetic to kinetic power (which, in shocks, can be thought of as an efficiency of magnetic field amplification), and $\eta$ is an efficiency factor defined in section~\ref{sec:shocks} which describes how close the diffusion is to the Bohm regime (equation~\ref{eq:diffusion}). 

In relativistic particle accelerators, the above expressions can be modified slightly to account for special relativistic effects. Ref.~\cite{achterberg_particle_2001} gives the Hillas energy in the form $E_H = \Gamma Ze B \beta R$, where $\Gamma$ is the bulk Lorentz factor of the shock, and $R$ and $B$ are given in the co-moving frame. Whether the particle really gains this Lorentz boost likely depends on the nature of the turbulence generated at the shock, and the details of the particle transport in the source region \cite{pelletier2008}. However, such a boost may potentially be important particularly if GRB internal shock or afterglow models are to reach rigidities of $10~{\rm EV}$ (see section~\ref{sec:grbs}). The power requirement can also include an additional $\Gamma^2$ factor \cite{lemoine_anisotropy_2009,rieger_active_2022}, although this cancels with the outflow opening angle, $\Theta$, if $\Theta \propto \Gamma^{-1}$. In any case, it is hard for a relativistic accelerator to reach optimal conditions with $\eta \approx 1$ (see  section~\ref{sec:shocks}), and so we take equations~\ref{eq:hillas} and \ref{eq:power} as our basic energetic requirements.

\subsection{Particle acceleration mechanisms}
Astrophysical fluids are often hot and ionized plasmas in which electrons and ions are unbound and free to move. The motion of these free charges tends to rapidly damp or screen any local electrostatic field present in the plasma. However, bulk, differential motions of the plasma, with a velocity $\boldsymbol{\vel}$, lead to a $-(\boldsymbol{\vel}/c) \times \boldsymbol{B}$ electric field which can accelerate particles, where the velocity can be thought of as the characteristic velocity of `scattering centres', in Hillas' language \cite{hillas_origin_1984}. Indeed, this electric field is the origin of the $\beta B$ term in the Hillas energy above. Rather than acceleration in some spark gap or monolithic electrostatic field, particles are thought to acquire nonthermal energies through interactions with magnetised plasma that lead to a so-called `Fermi' process: a gradual, stochastic acceleration in a $-(\boldsymbol{\vel}/c) \times \boldsymbol{B}$ electric field. 

 Fermi originally proposed that CRs gain energy from interactions with magnetised clouds \cite{fermi_origin_1949}, which, together with its derivatives, is now referred to as second-order Fermi acceleration because the fractional energy gain per scatter is proportional to $\beta^2$. In the late 1970s, a series of authors proposed first-order Fermi acceleration at shocks \citep{bell_acceleration_1978,blandford_particle_1978,axford_acceleration_1977,krymskii_regular_1977}, and since then Fermi processes have been extensively studied from various perspectives and are the subject of a number of review papers \citep[e.g.][]{drury_review_1983,blandford_particle_1987,matthews_particle_2020}. We refer the reader to these reviews for a detailed discussion. Here, we discuss some of the basic reasoning behind first-order Fermi processes and the physical mechanisms at work, as well as the astrophysical sites in which they can operate. 

\subsubsection{Shock Acceleration}
\label{sec:shocks}
The most famous example of first-order Fermi acceleration is shock acceleration. A shock is a converging flow, with $\nabla \cdot \boldsymbol{\vel} < 0$, and particles that cross the shock front gain a momentum boost proportional to the shock velocity $\beta$. The theory was laid out in the aforementioned series of papers, with Bell \cite{bell_acceleration_1978} providing a `microscopic', test-particle description, and Blandford \& Ostriker \cite{blandford_particle_1978} a `macroscopic' description using the Fokker-Planck equation. The basic result is that CRs crossing the shock front get an energy boost each time they do so, with a mean fractional energy gain
$\langle \Delta E/E\rangle = \beta$. At the same time, CRs are being swept away from the front at a rate which is also proportional to $\beta$. It is straightforward to show \cite[e.g.][]{bell_acceleration_1978,matthews_particle_2020} that the competition between these two effects -- energy gain, and escape -- leads to a power-law CR distribution of the form 
\begin{equation}
    \frac{dN}{dE} \propto E^{-q},
\end{equation}
with $q=2$ for the idealised example considered here. There are various effects that lead to a steepening of this spectral index, such as energy exchange with turbulent magnetic fields \cite{bell_steep_2019,diesing_spectrum_2019}, and in ultra-relativistic shocks an index of $q\approx 2.2-2.3$ is expected \cite{kirk_particle_2000,achterberg_particle_2001}.

By treating the crossing of the shock and the scattering by magnetic irregularities as a diffusive process with coefficient $D=\lambda c$, where $\lambda$ is the mean free path, it is possible to derive an acceleration time. In detail, this acceleration time should allow for different upstream and downstream diffusion coefficients \cite{lagage_cosmic-ray_1983}, but the basic form is 
\begin{equation}
    \tau_{\rm acc} \sim \frac{D}{\vel_s^2} \equiv \frac{\lambda}{\beta_s^2 c}.
    \label{eq:tau}
\end{equation}
The CR energy is maximised when the acceleration time is shortest, requiring small diffusion coefficients and large shock velocities (in the non-relativistic regime). The diffusion coefficient is often written in the form 
\begin{equation}
D \sim \eta r_g c
\label{eq:diffusion}
\end{equation}
where $\eta$ is the so-called gyrofactor; $\eta=1$ is the optimal Bohm regime where $\lambda \approx r_g$, and $\eta > 1$ otherwise leading to slower acceleration. It is easy to show that the Hillas energy is necessary but not sufficient by combining equations~\ref{eq:larmor}, \ref{eq:tau} and \ref{eq:diffusion}, and equating $\tau_{\rm acc}$ with $R/\vel_s$, giving the equation 
\begin{equation}
E_{\rm max} \sim \eta^{-1} Ze \beta B R.
\label{eq:bohm_hillas}
\end{equation}
Thus, the Hillas energy is only reached when $\eta = 1$ and Bohm diffusion applies, that is when $\lambda \approx r_g$. For this to happen, there must be strong turbulence with $\delta B / B \sim 1$ and structure in this turbulence on scales of the Larmor radius. These considerations show why the plasma physics of CR instabilites and the nonlinear, coupled acceleration process are important for understanding the maximum energy/rigidity attainable in a given accelerator. 

It was realised early on that CR-excited waves or MHD turbulence of some kind were needed to confine the CRs at the shock, allowing the CRs to cross many times and facilitate acceleration to high energies. Originally, Alfv{\'e}n waves driven by the resonant CR instability \citep{skilling_cosmic_1975} were invoked, but a new non-resonant or Bell instability was discovered \cite{lucek_non-linear_2000,bell_turbulent_2004}. The non-resonant instability has a number of advantages; it grows faster than the resonant instability and  creates turbulence on the scale of the Larmor radius of the particles driving the instability, providing a self-regulated process that allows acceleration to proceed close to the Bohm regime. The instability is thought to operate in supernova remnant (SNR) shocks where it is critical for providing the necessary magnetic field amplification, and growing the turbulent magnetic field to the Larmor radius of the highest energy particles. There is observational evidence that the Bohm regime is realised in SNR shocks \cite{stage_cosmic-ray_2006,uchiyama_extremely_2007}, suggesting they can get close to the special conditions needed for the Hillas energy to apply.  

In some sense it is natural to appeal to relativistic shocks as UHECR accelerators, given that some of the most powerful phenomena in the Universe involve ultrarelativistic outflows and invariably produce radiation from nonthermal electrons. The ultrarelativistic version of shock acceleration or first-order Fermi acceleration differs somewhat from its nonrelativistic counterpart. The expected spectral index is slightly steeper than the canonical shock acceleration value, with $q \approx 2.2-2.3$, the compression ratio is higher, the shock is quasi-perpendicular and significant anisotropies develop in the particle distribution function \cite{kirk_particle_2000,achterberg_particle_2001}. One might think the ultrarelativistic shocks are the most obvious sites for UHECR acceleration, since $\beta\to1$ maximises the Hillas energy, but a number of authors have shown that relativistic shocks have difficulties reaching ultrahigh energies \cite{pelletier2008,lemoine_electromagnetic_2010,reville_maximum_2014,bell_cosmic-ray_2018}. In particular, Ref. \cite{bell_cosmic-ray_2018} shows that the maximum energy is likely to be many orders of magnitude below EeV energies. This happens because the CR spectrum is steeper, so there is less energy to drive turbulence on UHECR Larmor radius scales, and the CRs also do not have {\em time} to drive large-scale turbulence, because they penetrate less far upstream and are quickly advected away downstream. There may be ways around these issues -- for example, if there is pre-existing turbulence in the upstream medium (see also Refs. \cite{huang_implications_2022,kirk_particle_2022} for relevant recent studies) -- but we will refer to these collective difficulties as the `relativistic shock problem' for UHECRs. 

\subsubsection{Magnetic Reconnection}
Magnetic reconnection -- the resistive dissipation of magnetic fields -- is another mechanism that can accelerate particles. In this case the transfer of energy is from magnetic energy to thermal and kinetic, a fraction of which can be passed on to nonthermal particles. Reconnection has received a lot of attention as a particle acceleration mechanism recently, for various reasons. The last decade has seen dramatic progress in using PIC (as well as test particle and hybrid MHD-PIC) simulations to study first 2D, and subsequently 3D, reconnection sites. A variety of particle acceleration mechanisms can operate in these reconnection sites; unlike shock acceleration there is a current sheet involved, and the electric field close to the reconnection X-point can inject particles or accelerate them to modest energies. After injection, Fermi mechanisms can take over. Various Fermi mechanisms and models have been proposed, with acceleration taking place by traversing the converging flows either side of the X-point \cite{de_gouveia_dal_pino_production_2005,giannios_uhecrs_2010,drury_first-order_2012} or within contracting plasmoids \cite{drake_electron_2006}. 

It is not yet clear how relevant magnetic reconnection is to the UHE, multi-EeV regime, though a number of authors have proposed it as a possible UHECR acceleration mechanism \citep{de_gouveia_dal_pino_ultra-high-energy_2000,giannios_uhecrs_2010,de_gouveia_dal_pino_ultra-high-energy_2021}. One potential difficulty is arranging for structure (and energy density) in the magnetic field to be present on a wide range of scales from the resistive scale up to the Larmor radii of UHECRs. In shock acceleration, the magnetic field is amplified and stretched via, e.g., the CR-driven non-resonant instability, but we (the authors) do not know of a convincing mechanism to arrange for such a magnetic field structure in reconnection sites at this stage. However, that does not mean one does not exist or will not be forthcoming in the future. 

\subsubsection{Other Mechanisms and General Comments}
As well as from shocks and reconnection, there are various other ways in which particles can gain large amounts of energy. Shear acceleration involves scattering across a shear layer \cite{rieger_particle_2019}, in a similar manner to shock acceleration, and has been proposed as a possible mechanism for UHECR acceleration at the edge of a relativistic jet. A detailed discussion of shear acceleration pertaining to UHECR acceleration in AGN jets is given by Rieger in a recent review \cite{rieger_active_2022}, who highlights some recent studies proposing one-shot or `espresso' acceleration in relativistic AGN jets \cite{caprioli_chemical_2017,kimura_ultrahigh-energy_2018,mbarek_bottom-up_2019}. Alternatively, shear acceleration can be rather gradual, and the details of the process depend on the thickness of the shear layer and the Kelvin-Helmholtz instabilities operating in the region \cite{rieger_microscopic_2006,rieger_active_2022,wang_particle_2022}. Various authors have also discussed second-order Fermi acceleration (Fermi II) by MHD turbulence in, for example, giant radio lobes \cite[][see also section~\ref{sec:agn}]{hardcastle_high-energy_2009,osullivan_stochastic_2009,wykes_mass_2013}. Finally, there is the possibility of acceleration by `unipolar induction', whereby a rapidly rotating magnetic field in, e.g., a pulsar magnetosphere generates a large potential difference \cite{goldreich_pulsar_1969,blandford_acceleration_2000,blasi_ultra-high-energy_2000,fang_newly_2012}. We do not provide a detailed account of these three processes (shear, Fermi II, unipolar induction), and neither is this an exhaustive list, but we we do touch on some of them in more detail, with the relevant astrophysical context, in section~\ref{sec:where}. In our discussions hereafter, we will try to keep the arguments general, based on the basic energetics and physical conditions of the system, but a bias in emphasis towards shock acceleration is probably inevitable given the background of the authors, and the relative maturity of each mechanism at the time of writing. 

\subsection{UHECR Losses and Escape}
\label{sec:losses2}

The Hillas energy and power requirements above and the maximum energy derived from shock acceleration apply when the CR maximum energy is limited by either the escape time or dynamical time. In sources with strong magnetic fields or intense radiation fields, losses can instead limit the maximum energy. Synchrotron losses for CR nuclei with relative atomic mass $A$ occur on a timescale $\tau_{{\rm sync}} = 142~{\rm yr}~(A/Z)^4~E_{{\rm EeV}}^{-1} B^{-2}$, where $E_{\rm EeV}$ is the energy in EeV. By equating this with the acceleration time (equation~\ref{eq:tau}), we can write the maximum energy in a magnetic field of strength $B$ as
\begin{equation}
E_{\rm max,sync} = 200~{\rm EeV}~\left(\frac{1}{\eta Z^3}\right)^{1/2}~\left(\frac{B}{{\rm G}}\right)^{-1/2}~\beta A^2,
\end{equation}
which can be restrictive even for optimum acceleration conditions in strong magnetic field sources. Equivalently, we can invert this equation to write a maximum magnetic field strength for acceleration to a given energy,
\begin{equation}
B_{\rm max} = 400~{\rm G}~\left(\frac{1}{\eta Z^3}\right)~\left(\frac{E}{10~{\rm EeV}}\right)^{-2}~\beta^2 A^4 ,
\label{eq:bmax}
\end{equation}
which illustrates the challenge of accelerating UHECRs in highly magnetised environments as discussed by various authors \cite[][see section~\ref{sec:where} for source implications]{aharonian_constraints_2002,murase_high-energy_2008,wang_origin_2008,peer_radio_2009,globus_uhecr_2015,piro_ultrahigh-energy_2016}. Interactions with ambient radiation fields during both the acceleration and escape of CRs can also limit the maximum energy through the same processes described in section~\ref{sec:losses}, although the details depend on the radiation field shape and intensity considered. Considering photopion losses from protons, Ref. \cite{aharonian_constraints_2002} derives an approximate limit on the source radiative luminosity at a distance $R$ from the source given by $L_{\gamma}<5\times 10^{44}~{\rm erg~s}^{-1}~\bar{\epsilon}~(R/10^{17}{\rm cm})$
where $\bar{\epsilon}$ is the energy of the maximum of the integral photon spectrum in eV. Such an upper limit might seem counter-intuitive given that we also found a lower limit on power from equation~\ref{eq:power}, but the latter is a magnetic power limit as opposed to a radiative one. The acceleration of UHECRs therefore favours sources which are neither too {\em radiatively efficient} nor too {\em compact} -- ideally we need large amounts of energy to be dissipated so that a large product $\beta BR$ can be maintained, but without the energy densities in magnetic fields or radiation fields becoming too large. 

We close this section by noting that the losses within, and escape from, the acceleration site and immediate environment can have interesting implications for the emergent spectrum and composition of the UHECRs. For example, in the Unger-Farrar-Anchordoqui model \cite{unger_origin_2015}, photodisintegration in the source environment can naturally reproduce the location of the UHECR ankle, shape of the UHECR spectrum and composition trends (see also Ref.~\cite{globus_complete_2015}). Similarly, diffusive escape modifies the spectral shape below a critical energy at which the escape time, $\tau_{\rm esc}$, is equal to the source age \cite[e.g.][]{kotera_propagation_2009,matthews_particle_2021}. Above this energy, the at-source spectrum is gradually recovered. Furthermore, high rigidity CRs escape more quickly than low rigidity CRs, so the escaping UHECRs can be lighter than the internal CRs, although the details depend on a complex interplay between the source activity, and the cooling/escape timescales \cite{matthews_particle_2021}. 

\subsection{UHECR Source Checklist}
With the above arguments in mind, and referring the reader to the references given for greater detail, we propose that the following criteria form a basic `checklist' that a source or source population must satisfy to be a realistic UHECR candidate:
\begin{itemize}
    \item The source must have a large product $\beta B R$ to satisfy the Hillas condition (equation~\ref{eq:hillas})
    \item The source must dissipate a large amount of power in (for example) a shock or a site of magnetic reconnection (equation~\ref{eq:power})
    \item If the acceleration is diffusive, the diffusion coefficient must approach the Bohm regime or near-optimal conditions (e.g. equation~\ref{eq:bohm_hillas}) across a range of energies
    \item If the acceleration is at a shock, the shock probably cannot be highly relativistic 
    \item The CRs must not undergo restrictive losses due to, e.g. curvature radiation, synchrotron radiation, adiabatic expansion, or interactions with photons
    \item The source must be within a composition- and energy-dependent horizon from the Earth, or produce UHECRs with such efficacy that a substantial UHECR luminosity still reaches us. 
    \item The source must be common and powerful enough to produce the observed UHECR flux.
\end{itemize}
Meeting all of these criteria turns out to be a challenge for any astrophysical source. However, it is also difficult to assess the relative merit of the sources given that (i) the underlying physics is complex and far from settled, and (ii) many of the estimates of velocities, magnetic field strengths, and jet/outflow powers are subject to large astrophysical uncertainties. Nevertheless, we will soldier on and discuss possible UHECR sources with this list of requirements as our guide. 

\section{Astrophysical Sources of UHECRs ({\em Where?})}
\label{sec:where}
The detections of anisotropies in UHECR data from the {\sl Pierre Auger Observatory} (PAO) and {\sl Telescope Array} mean that we are entering an exciting era for UHECR astrophysics. In this section, we will first look at anisotropy data to see what the data alone tell us, before discussing the overall prospects of a host of astrophysical candidates.

\subsection{UHECR anisotropies}
Detecting statistically significant anisotropies in the arrival directions of UHECRs is a key goal of TA and PAO, and is an essential step for uncovering the origin of UHECRs. In 2017, PAO reported a large-scale anisotropy in the arrival directions of 30,000 CRs above $8$~EeV at $5.2\sigma$ significance. The anisotropy is well-described by a dipole with $6.5\%$ anisotropy. TA has also undertaken large-scale anisotropy searches \cite{abbasi_search_2020}, with results that are consistent with both isotropy and the PAO dipole. The detection of a significant UHECR dipole is a spectacular and important result. It more-or-less confirms some aspects of UHECR origins which had long been suspected: that UHECRs are extragalactic in origin, and are not isotropic. However, it is extremely difficult to pinpoint UHECR sources from a large-scale dipole on the sky. Moving to higher energies results in a trade-off. On the one hand, higher energy typically means higher rigidity, resulting in smaller magnetic deflections and anisotropies that emerge on smaller angular scales. These anisotropies can feasibly be correlated with astrophysical source catalogues. On the other, the statistics drop off markedly and so the number of events one is able to analyse can become prohibitively small. 

\begin{figure}
\centering
\includegraphics[width=0.9\linewidth]{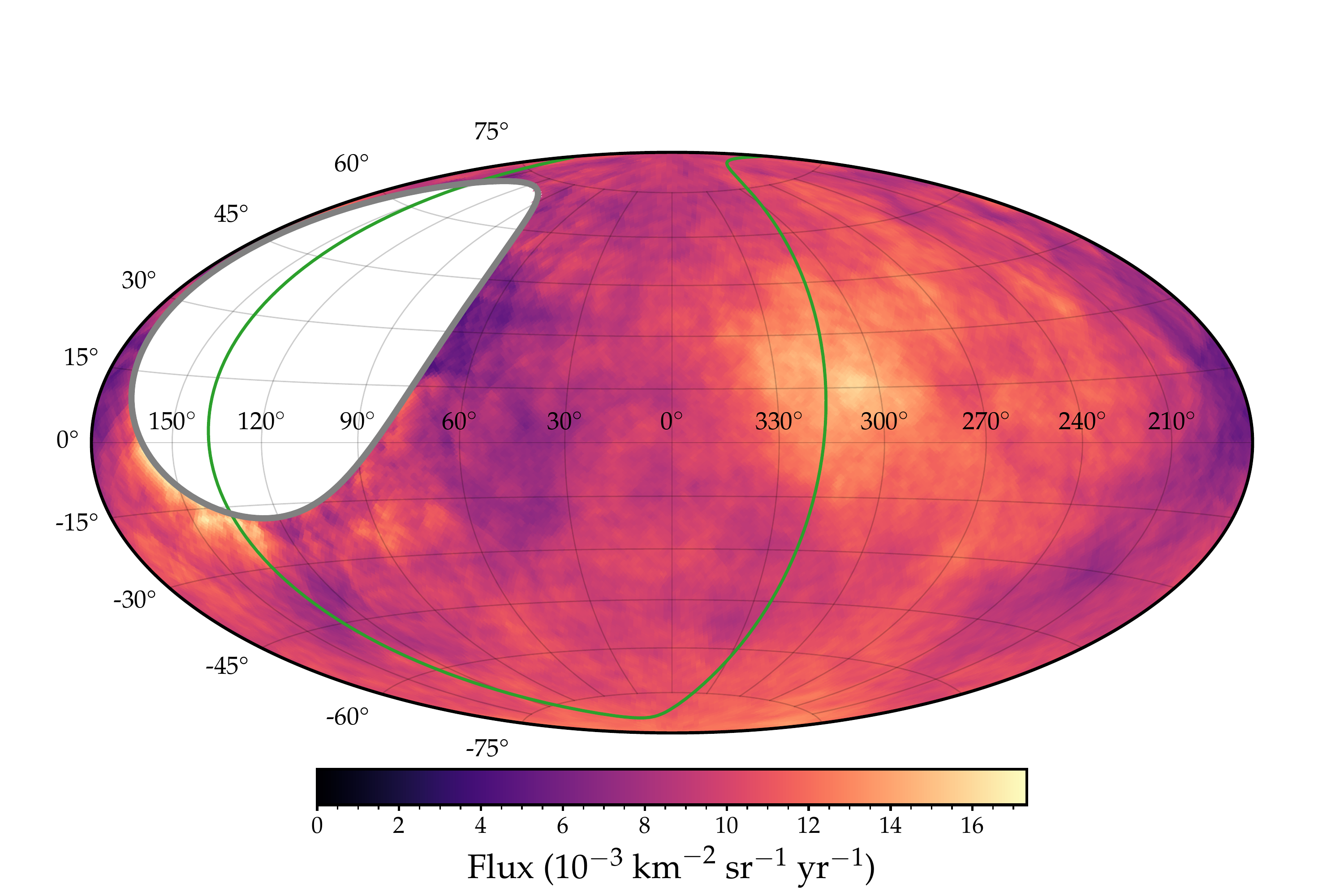}
\caption{
UHECR flux map, in Galactic coordinates, from PAO above $41$~EeV comprising 1274 events, produced from the data made publicly available by PAO \cite{pierre_auger_collaboration_arrival_2022}. A top-hat smoothing of radius $25^\circ$ has been applied. The supergalactic plane and PAO exclusion are marked in green and grey, respectively.  
}
\label{fig:auger_flux}
\end{figure}

Both TA and PAO have reported indications of anisotropy on intermediate angular scales ($\approx5^{\circ}-25^{\circ}$ search radii) at $\gtrsim 40~{\rm EeV}$ energies. PAO found (model-dependent) correlations with catalogues of star-forming galaxies ($4\sigma$) and AGN ($3.2\sigma$) \cite{aab_indication_2018}, with anisotropic fraction of around $5-10\%$. Here a `UHECR luminosity proxy' must be adopted and PAO used gamma-ray and radio fluxes as weights for the relative luminosity of each source in UHECRs. In a more recent study \cite{pierre_auger_collaboration_arrival_2022}, PAO presented a detailed investigation of 2635 events with reconstructed energy $>32~{\rm EeV}$. The arrival directions from this study are shown in Fig.~\ref{fig:auger_flux}. An excess in flux can be seen just above the Galactic plane, approximately in the direction of Centaurus, with a hint of additional excesses at southern Galactic latitudes and just below the PAO exclusion area. Using updated catalogues for gamma-ray luminosities as well as considering other catalogues with different luminosity proxies, Ref. \cite{pierre_auger_collaboration_arrival_2022} found post-trial $p$-values of $<10^{-3}$ for catalogues comprising jetted AGN, galaxies traced by near-infrared emission, X-ray AGN and star-forming galaxies. Each of these searches resulted in top-hat search radii in the range $22^{\circ}-25^{\circ}$ and threshold energies of $38-40~{\rm EeV}$, and the most significant correlation was with the star-forming galaxies traced by their radio emission, giving $p=3.2\times 10^{-5}$ corresponding to $>4\sigma$ significance. The same study also conducted a series of less model-dependent searches for correlations with structures such as the supergalactic and Galactic plane; although none of these were statistically significant, they did find a $>4\sigma$ correlation with the Centaurus region on the sky, which is responsible for driving much of the correlation seen in the catalogue searches (since Cen A, NGC 4945 and M83 all lie in this area). More details on this study can be found in the contribution of C. Galleli, on behalf of PAO, to these proceedings.

TA have also reported anisotropies on intermediate angular scales, with a particular excess that is often referred to as the `TA hotspot' \cite{abbasi_indications_2014}. TA reported an excess of events above $57~{\rm EeV}$ fairly close to the supergalactic plane, at ${\rm RA}=146.7^\circ$, ${\rm Dec.}=43^\circ$, which is roughly the direction of M82. Originally, the post-trial (local Li-Ma) significances was $3.4\sigma$ ($5.1\sigma)$, while more recent updates put the post-trial significance at $2.9-3.2\sigma$ \cite{kawata_ta_2019,sagawa_highlights_2022}. TA have also reported another excess at slightly lower energies, in the approximate direction of the Perseus-Pisces supercluster \cite{telescope_array_collaboration_indications_2021}. 

Finally, in addition to the TA and PAO searches, joint PAO and TA efforts have been undertaken to build full-sky maps of UHECR arrival directions by correcting for the different exposures and systematics between the two experiments \cite{biteau_covering_2019,di_matteo_full-sky_2020}. The results emerging from this show excesses roughly correlated with the supergalactic plane (or local sheet/CoG, which follows a similar path on the sky), with particular hotspots in the direction of NGC 253/Fornax, Cen A/NGC2945, and M82. Although the statistical significance of any correlations with these local planar structures is still fairly marginal ($\sim 3\sigma$) \cite{di_matteo_full-sky_2020}, the full-sky UHECR arrival directions seem to highlight the importance of nearby sources and/or structures, especially when considered in tandem with the arguments for local sources discussed in section~\ref{sec:losses}. 

\subsection{Possible Source Classes}
At this stage, unambiguous source identifications from anisotropy data alone are not possible. We must therefore consider carefully the underlying astrophysics when considering the best candidate UHECR sources. Here, we will make a whistle-stop tour of possible sources based on the physics underpinning particle acceleration to ultrahigh energies discussed in the previous section. 

\subsubsection{Star-forming Galaxies}
\label{sec:star}
Particle acceleration in star-forming galaxies has been extensively studied, but the indication of anisotropy reported by PAO \cite{aab_indication_2018}, with a $4\sigma$ correlation with a `starburst' catalogue, ignited interest in this class of objects as UHECR sources. We prefer to use the term `star-forming' rather than `starburst' for this catalogue, as many of the galaxies have fairly typical star formation rates rather than the extreme values normally associated with starbursts. A number of local star-forming galaxies are known to be gamma-ray emitters and include {\sl bona fide} starbursts such as M82 \cite{abdo_detection_2010,ackermann_gev_2012}. It is therefore clear that high-energy particle acceleration does take place in these sources.

A number of authors have either discussed the physics of particle acceleration in these starburst `superwinds' or proposed them as UHECR sources \cite{romero_particle_2018,anchordoqui_acceleration_2018,peretti_particle_2022,anchordoqui_deciphering_2022}. The superwinds are thought to be caused by dramatic bursts of star formation, with the combined effect of supernovae and stellar winds (and possibly also CR pressure from low energy CRs) driving a powerful kpc-scale outflow \cite{heckman_nature_1990,strickland_starburst-driven_2000}. The kinetic powers are estimated at $\sim10^{42}~{\rm erg~s}^{-1}$ for sources like NGC~253 and M82, with shock velocities of $\sim 1000~{\rm km~s}^{-1}$ \cite{strickland_starburst-driven_2000,strickland_supernova_2009,romero_particle_2018}. Taking these numbers as characteristic we find a maximum rigidity estimate of 
\begin{equation}
{\cal R}_{\rm max} \sim 
0.15~{\rm EV}~\left[
\left(\frac{\epsilon_b}{0.1}\right)
\left( \frac{Q_k}{3\times10^{42}~{\rm erg~s}^{-1}} \right)
 \left( \frac{\vel}{1000~{\rm km~s}^{-1}} \right)
\right] ^{1/2} \eta^{-1}
\end{equation}
for particles accelerated by starburst superwinds, once again emphasizing the importance of the electric field and associated velocity term. This estimate is on the optimistic end of the more detailed calculations presented by Ref. \cite{romero_particle_2018}, and would require quite efficient magnetic field amplification. UHECRs at our target maximum rigidity of $10~{\rm EV}$ would therefore appear to be beyond the capabilities of starburst superwinds. Star-forming galaxies may still be the sources of UHECRs on the sky through their accumulated populations or historical record of magnetised neutron stars (magnetars and/or pulsars), gamma-ray bursts or tidal disruption events. We discuss these source classes in sections~\ref{sec:pulsars}, \ref{sec:grbs} and \ref{sec:others}, respectively. In addition, star-forming galaxies are polluters of the circumgalactic medium \cite[CGM; e.g.][]{tumlinson_large_2011,lopez-rodriguez_strength_2021}, a process which could produce magnetic fields on large scales and act as barrier to, or reflector of, UHECRs (see section~\ref{sec:when}). 

\subsubsection{Radio Galaxies and AGN}
\label{sec:agn}
Radio galaxies (using our adopted definition) are active galaxies which produce giant, kpc-scale jets that emit synchrotron radiation in the radio band \cite{hardcastle_radio_2020}. They have been commonly suggested as UHECR sources \cite{hillas_origin_1984,rachen_extragalactic_1993,norman_origin_1995,dermer_ultra-high-energy_2009,massaglia_role_2007,wykes_mass_2013,wykes_uhecr_2017,eichmann_ultra-high-energy_2018,eichmann_summing_2019}, and the local radio galaxy Centaurus A (Cen A) is a particularly compelling candidate \cite{hardcastle_high-energy_2009,biermann_centaurus_2012,sahu_hadronic-origin_2012}. Radio galaxies are known to accelerate non-thermal particles as revealed by their radio, X-ray and gamma-ray emission, but the particle acceleration mechanism and sites of energy dissipation vary from source-to-source. Their radio morphology gives important clues as to their particle acceleration physics, and is broadly split into two Fanaroff-Riley (FR) classes \cite{fanaroff_morphology_1974}: FR-I sources are brighter in the centre and resemble a disrupted plume of jet material, whereas FR-II sources are brightest far from the nucleus and remain well-collimated until they reach the termination shock, producing a radio hotspot. The FR dichotomy is thought to be caused by a combination of jet power and environment \cite{hardcastle_radio_2020}, with FR-II sources generally being associated with powerful sources and/or poorer group or cluster environments (although Ref. \cite{mingo_revisiting_2019} has shown FR-II morphologies can be produced down to rather low radio luminosities). From a particle acceleration perspective, it is generally thought that FR-II sources primarily accelerate the synchrotron-emitting electrons in the hotspot associated with the jet termination shock \cite[e.g.][]{meisenheimer_particle_1986,meisenheimer_synchrotron_1989,araudo_particle_2015} with additional particle acceleration along the jet itself \cite{perley_jet_1984,gentry_optical_2014,hardcastle_deep_2016} and in the lobes or backflows \cite{wykes_mass_2013,matthews_ultrahigh_2019,mukherjee_new_2020}. By contrast, many FR-I sources appear to continuously accelerate nonthermal electrons along their length in a distributed, in-situ process \cite{laing_spectra_2013,2020Natur.582..356H}, although they can also have features such as knots \cite{bridle_deep_1994,goodger_long-term_2010} or bow shocks \cite{croston_high-energy_2009,timmerman_origin_2021}. 

Given their higher average jet powers and strong termination shocks, FR-II radio galaxies are the natural class of UHECR accelerators from an energetic perspective \cite{hillas_origin_1984,rachen_extragalactic_1993,norman_origin_1995}. The maximum energy in radio hotspots may be severely limited due to their relativistic nature \cite{araudo_maximum_2018}, which motivated Matthews et al. to propose a model in which UHECRs are accelerated in multiple shocks of $\beta \approx 2$ in supersonic backflows. The maximum rigidity in these backflows is estimated, using hydrodynamic simulations, at $50~EV$. However, powerful radio galaxies are also relatively rare, with only a handful of FR-II sources within a few hundred Mpc \cite{massaglia_radio_2007,matthews_fornax_2018}. This has lead various authors to consider local FR-I radio galaxies such as Cen A and Fornax A, which also have compelling associations with UHECR anisotropies \cite{matthews_fornax_2018}; however, in Cen A the current jet power is thought to be $\sim10^{43}~{\rm erg~s}^{-1}$ \cite{wykes_mass_2013} and the bow-shock associated with the current activity is not fast or powerful to accelerate UHECRs \cite{croston_high-energy_2009}. 
Thus, although particle acceleration in mildly or non-relativistic shocks associated with radio galaxies seems quite attractive as an UHECR origins story, this requires flickering-type variability or jet powers which were significantly higher in the past \citep{matthews_fornax_2018,matthews_cosmic_2019,matthews_particle_2021}. A further challenge is to make sure that, if the jet is initially leptonic, then it entrains a significant hadronic component \cite{wykes_mass_2013,wykes_uhecr_2017}. Alternatively, Hardcastle, Wykes and collaborators \cite{hardcastle_high-energy_2009,wykes_mass_2013,wykes_uhecr_2017} have suggested Fermi II acceleration in Cen A's turbulent lobes, which could produce UHECRs for very fast Alfv{\'e}n speeds. 
Furthermore, Eichmann and collaborators have shown that the UHECR spectrum can be reproduced from a population of radio galaxies including local sources like Cen and Fornax A, even when taking into account the detailed source physics and propagation \cite{eichmann_ultra-high-energy_2018,eichmann_summing_2019,eichmann_high_2019}.

Finally, AGN winds, which are responsible for X-ray `ultra-fast outflow' features \cite{gofford_suzaku_2013} and for quasar blue-shifted broad absorption lines \cite{weymann_broad_1985,rankine_bal_2020}, can drive shocks into their surroundings. These shocks may accelerate particles and produce synchrotron and inverse Compton emission \cite{nims_observational_2015}. Although there is no definitive detection, there are observational hints of quasar wind contributions to radio and gamma-ray emission \cite{ajello_gamma_2021,rankine_placing_2021,richards_probing_2021,petley_connecting_2022}. Quasar winds might reach $Q_k \sim 5 \times 10^{44}~{\rm erg~s}^{-1}$ \cite{nims_observational_2015}, if the outflow is $5\%$ efficient compared to the bolometric power. The maximum shock velocity at the reverse shock is $\lesssim0.1c$, suggesting a maximum rigidity of a few EeV from equation~\ref{eq:power}. For estimates based on Ref.~\cite{nims_observational_2015} of $B\sim 1$~mG and $R\sim 0.1$~kpc, we obtain ${\cal R}_H \sim 10~{\rm EV}$. AGN winds were investigated as UHECR sources by Ref.~\cite{wang_ultrahigh_2017}, who found maximum proton energies of $\sim 100~{\rm EeV}$ for optimistic parameters. Thus, AGN winds merit further investigation, but it is hard to draw any firm conclusions here, and the current lack of evidence for nonthermal particles makes them less compelling as UHECR accelerators. 

\subsubsection{Gamma-ray bursts}
\label{sec:grbs}
Gamma-ray bursts (GRBs) are violent, catastrophic explosions that release flashes of gamma-rays and prodigious amounts of energy. The general paradigm \cite[e.g.][]{meszaros_gamma-ray_2019} is that short GRBs are associated with merging neutron stars, and long GRBs with `collapsars', in which a massive stellar core collapses to form a black hole; the latter are more common and so probably more interesting as UHECR sources. GRBs are known to accelerate particles, most likely in shocks, with the prompt emission thought to be produced by highly relativistic colliding internal shocks, giving way to a longer-lived afterglow phase powered by a forward shock. Additionally, emission from the reverse shock can be detected at early times \cite{laskar_reverse_2016}. MAGIC has detected early time afterglow emission at hundreds of GeV coincident with GRB~190114C \cite{acciari_teraelectronvolt_2019}, and HESS detected late time afterglow emission up to 3~TeV from GRB~190829A \cite{2021Sci...372.1081H}. Furthermore, during the writing of these proceedings, LHAASO has reported an exciting detection of early time emission up to $18$~TeV associated with the brightest (but not most intrinsically luminous) GRB to-date, GRB~221009A \cite{huang_lhaaso_2022}. GRBs are clearly excellent particle accelerators up to at least the TeV regime -- but can they reach ultrahigh energies? 

GRBs have been regularly discussed as UHECR sources, since a few years after workable theoretical models for the GRB engine and resulting fireball emerged \cite{waxman_cosmological_1995,vietri_acceleration_1995,waxman_high-energy_2004}. They are attractive on account of their energetics; a GRB releases a total isotropic energy of up to $E_{\rm iso} \sim 10^{54}~{\rm erg}$ \cite[e.g.][]{perley_afterglow_2014} in a relatively short amount of time, and a large fraction of this power is dissipated through a strong shock. Waxman \cite{waxman_cosmological_1995} and Vietri \cite{vietri_acceleration_1995}
argued that protons could reach energies in excess of $10^{20}~{\rm eV}$ and that the observed UHECR flux above $10^{20}~{\rm eV}$ could be explained as long as GRBs release similar amounts of energy in $>10^{20}~{\rm eV}$ CRs to the amount released in gamma-rays. More recent studies have confirmed the need for rather efficient UHECR production, as well as high baryon loading, to produce the UHECR flux at Earth \cite{waxman_high-energy_2004,globus_uhecr_2015,baerwald_are_2015}. However, if these efficiencies can be accommodated, then GRB models can in principle reproduce the observed UHECR spectrum rather well \cite{globus_uhecr_2015}. 

As mentioned in section~\ref{sec:shocks}, an additional difficulty with UHECR acceleration in GRB blast waves arises from the ultra-relativistic nature of the shock. Even GRB afterglows have initial shock Lorentz factors of $\sim 100$, approximately evolving with observer time, $t_{\rm obs}$ as $\Gamma_{\rm sh} \propto t_{\rm obs}^{-3/8}$ in the case of a uniform density medium \cite{blandford_fluid_1976,piran_physics_2004}. This evolution leads to a relativistic-Newtonian transition around a year after the initial burst \cite{piran_physics_2004}, and this non-relativistic phase dissipates significantly less power through the shock. For GRB afterglow shocks to accelerate UHECRs, one therefore either needs to circumvent the relativistic shock problem, or find another way to reach the UHE regime (e.g. by reconnection or shear acceleration). While in some sense this is the same problem discussed above for radio galaxies, in radio galaxies the jet bulk Lorentz factors are thought to be much lower and the presence of mildly or non-relativistic shocks in backflows and, perhaps in some cases, jets themselves, makes them more conducive to UHECR acceleration than GRBs. Further work is clearly needed to better understand the particle acceleration and shock physics in GRBs, but in principle the high energy leptonic emission is able to provide a useful probe of the turbulent plasma conditions. This was demonstrated recently by Ref. \cite{huang_implications_2022}, with the recent TeV gamma-ray emission detected perhaps already challenging our understanding of the nature of turbulence in relativistic shocks.

\subsubsection{Magnetised Neutron Stars}
\label{sec:pulsars}
A highly magnetised, rapidly rotating neutron star can accelerate particles with a voltage drop of $\phi =\Omega^2 \mu/c^2$,
where $\Omega$ is the angular velocity and $\mu = R_{\rm NS}^3 B$ is the magnetic moment. This voltage is time-dependent since the NS can spin down (or up) and thus $\Omega$ can change. For sufficiently small periods and large magnetic dipole moments, the maximum energy is $E_{\rm max} \sim 10^{21}~{\rm eV}~Z~\mu_{33}~P_{\rm ms}^{-2}$ \cite[e.g.][]{piro_ultrahigh-energy_2016}, where $\mu_{33}$ is the magnetic dipole moment in units of $10^{33}~{\rm G~cm^{3}}$, typical for newly born magnetars with $B\sim 10^{15}~{\rm G}$ \cite{mus_first_2019}, and $P_{\rm ms}$ is the orbital period in milliseconds. For young pulsars with $B\sim 10^{12}~{\rm G}$ the maximum energy is a factor of 1000 lower. For magnetars, this characteristic energy is easily within the UHECR regime for reasonable parameters; however, the problem with this mechanism is that particles accelerated in the NS magnetosphere (inside the light cylinder) undergo debilitating curvature losses, while CRs also undergo rapid synchrotron losses in such strong magnetic fields. Furthermore, the strong magnetic fields can be screened by prolific pair production \cite{goldreich_pulsar_1969,blandford_acceleration_2000}. These limitations have instead led various authors to consider acceleration in the pulsar wind termination shock or nebula associated with a very young, rapidly spinning pulsar \cite{blasi_ultra-high-energy_2000,fang_newly_2012,lemoine_ultra-high_2015} or magnetar \cite{arons_magnetars_2003,piro_ultrahigh-energy_2016}, or from $\nabla B$ drifts in the wind nebula \cite{bell_cosmic_1992}. The termination shocks are likely to be highly relativistic and so the same issues discussed above for radio galaxies and GRBs apply. However, the Crab nebula does seem to be able to accelerate particles in close to optimal conditions; LHAASO have detected PeV emission from the Crab nebula \cite{lhaaso_collaboration_petaelectron_2021} that requires $\eta^{-1} = 0.15$ in our notation. It seems fairly likely that Crab-like PWNe are perhaps PeVatrons, but probably not UHECR sources, with a maximum proton energy of $\sim 30~{\rm PeV}$ \cite{blandford_acceleration_2000}. The higher spin-down luminosities of rapidly spinning magnetars make them more attractive, and the winds they power are feasible sites of particle acceleration to rigidities of $\sim 10~{\rm EV}$ \cite{piro_ultrahigh-energy_2016}. 

\subsubsection{Other Objects}
\label{sec:others}
A veritable menagerie of other astrophysical objects have been proposed as UHECR sources; we only briefly discuss tidal disruption events (TDEs) and clusters of galaxies. TDEs are transient events that occur when a star passes close to a supermassive black hole and is ripped apart and accreted \cite{gezari_tidal_2021}. Radio emission from jets in TDEs has been detected in a handful of sources \cite{burrows_relativistic_2011,de_colle_jets_2020}, and these jets could be interesting as sites of UHECR acceleration providing the CRs can survive in the intense radiation field and the relativistic shock problem can be overcome. Ref.~\cite{zhang_high-energy_2017} finds the former should be possible in the reverse shock and estimates a maximum energy of $64~{\rm EeV}$ for protons (see also Ref. \cite{guepin_ultra-high-energy_2018}). Cluster shocks -- associated with gas accretion or galaxy mergers --  are potential sites of UHECR acceleration, because they are extremely large (virial radii $\sim 1~{\rm Mpc}$). The intracluster medium (ICM) magnetic field is $\sim 1~{\rm \mu G}$ and the shock velocity is $\sim 1000~{\rm km~s}^{-1}$, leading to a Hillas rigidity of $R_{\rm H}\sim 3 ~{\rm EV}$. This estimate is roughly in line with the maximum proton energy suggested by Kang for acceleration by multiple weak shocks in the ICM \cite{kang_diffusive_2021}, and the calculation by Ref. \cite{ptuskin_acceleration_2009}, who find cluster accretion shocks can produce UHECRs from $\sim 2-10~{\rm EeV}$ but struggle to produce the very highest energies. However, Ref.~\cite{kang_cluster_1996} show that UHECRs with energies up to $\sim 60~{\rm EeV}$ could be accelerated in larger $(R \sim 5~{\rm Mpc})$ accretion shocks associated with `caustics' if Bohm diffusion applies, finding that the Virgo cluster could make a substantial contribution to the observed UHECR flux at $\sim 30~{\rm EeV}$.

\begin{figure}[ht]
\centering
\includegraphics[width=1.0\linewidth, clip=True, trim=0 0.8in 0 0]{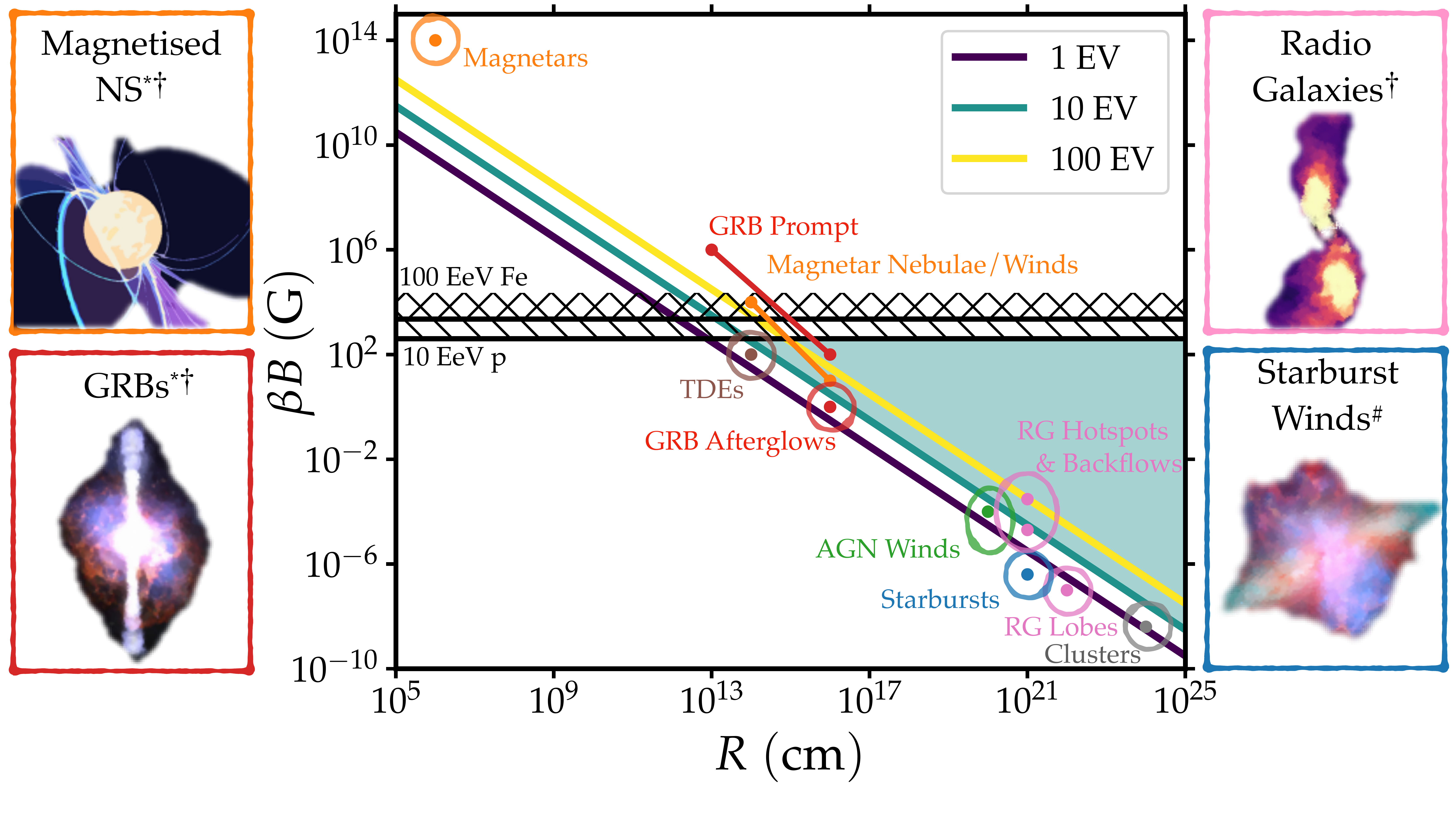}
\caption{
Modified Hillas plot showing the maximum electric field available, $\beta B$, plotted against characteristic size. The three diagonal lines show parameters needed to achieve the labelled maximum rigidity. The coloured points show the values from table~\ref{hillas_table}, the circles give a feel for the uncertainties, and points connected by lines have multiple estimates. The symbols ($\dagger,*,\#$) represent caveats; the meanings of $\dagger$ and $*$ are defined in the text, while $\#$ means the source does not dissipate sufficient kinetic or magnetic power (equation~\ref{eq:power}). The lines with hashed regions mark the maximum magnetic field for acceleration of 10 EeV protons and 100 EeV Fe nuclei (equation~\ref{eq:bmax}). Broadly speaking, a vaiable UHECR source must sit in the region that is shaded in translucent green. 
}
\label{fig:mod_hillas}
\end{figure}

\subsection{A Modified Hillas Diagram}
In Hillas' 1984 work, a number of useful diagnostic plots for identifiying UHECR sources were provided. Figure 1 of the same paper is often referred to as a `Hillas diagram', in which characteristic size is plotted along the $x$-axis, and magnetic field along the $y$-axis. One can then obtain a rough estimate of the feasibility of UHECR sources by drawing diagonal lines of constant rigidity, an exercise that has been regularly used to inform discussions of UHECR origins \cite[e.g.][]{hillas_origin_1984,ptitsyna_physics_2010,alves_batista_open_2019}. However, technically the diagram only conveys the ability of a source to confine UHECRs of a given rigidity -- it does not include the characteristic velocity of the accelerator and thus does not capture the impact of the electric field, ${\cal E} \propto \beta B$. The velocity of the scatterers {\em is} plotted in Hillas' figure 6, which is perhaps more informative in terms of which sources can actually accelerate to the UHE regime. Here we instead show a `modified Hillas diagram' in Fig.~\ref{fig:mod_hillas}, in which the electric field available, $\beta B$, is made explicit and plotted directly on the $y$-axis. The overall result is rather similar to the usual diagram, except that some sources move downwards in the parameter space (if they have $\beta<1$). The main source classes discussed in the text have been given extra prominence in their labelling and the values used in our estimates are given in table~\ref{hillas_table}. The plot does not capture the detailed physics of UHECR acceleration, or the relativistic shock problem, and nor does it contain information about the number density or luminosity density of sources; nevertheless, this modified Hillas diagram acts as a useful summary of the maximum rigidities attainable in UHECR candidate sources, while emphasizing the need for more accurate estimates of $\beta$ and $B$ in cosmic accelerators.  

\begin{table}[ht]
\centering
\begin{tabular}{cccccc}
\hline
Source Class &  $\beta$ & $R~({\rm cm})$ & $B~({\rm G})$ & Ref. & Section \\ \hline \hline
Magnetars$^{\dagger,*}$   &   $1$         &         $10^{6}$       &     $10^{14}$    &\cite{piro_ultrahigh-energy_2016} & \ref{sec:pulsars} \\
Magnetar Winds$^{\dagger,*}$   &  $1$         &       $10^{14},10^{16}$       &    $10^{4},10$    & \cite{piro_ultrahigh-energy_2016} & \ref{sec:pulsars} \\
GRB Prompt/Internal$^{\dagger,*}$       &     $1$   &    $10^{13},10^{16}$                         &  $10^6,100$ & \cite{piran_magnetic_2005,globus_uhecr_2015}  & \ref{sec:grbs} \\
GRB Afterglows$^{\dagger}$       &     $1$   &        $10^{16}$                  &   $1$        &             \cite{piran_magnetic_2005}   & \ref{sec:grbs}                      \\
RG Hotspots$^\dagger$    &    $1$     &       $10^{21}$        &      $3\times10^{-4}$       &   \cite{araudo_maximum_2018} & \ref{sec:agn}\\  
RG Backflows             &    $0.2$     &       $10^{21}$        &     $10^{-4}$     &  \cite{matthews_ultrahigh_2019}& \ref{sec:agn} \\  
RG Lobes                 &    $0.01$     &       $10^{22}$        &     $10^{-5}$       & \cite{croston_x-ray_2005} & \ref{sec:agn}\\  
Starburst Winds          &    $0.004$     &      $10^{21}$      &    $10^{-4}$      &    \cite{strickland_starburst-driven_2000,romero_particle_2018} & \ref{sec:star} \\    
TDE Jets$^{\dagger}$     &     $1$    &        $10^{14}$        &     $100$       &  \cite{zhang_high-energy_2017}  & \ref{sec:others} \\ 
Cluster Shocks             &   $0.004$    &       $10^{24}$        &    $10^{-6}$        &  \cite{kang_cluster_1996,ptuskin_acceleration_2009}  & \ref{sec:others} \\ 
AGN Winds          &    $0.1$     &     $10^{20}$          &      $10^{-3}$     & \cite{nims_observational_2015} &  \ref{sec:agn} \\ 
\hline
\end{tabular}
\caption{Table of characteristic values of parameters adopted for the modified Hillas diagram (Fig.~\ref{fig:mod_hillas}). In some cases multiple values are quoted designed to crudely represent the range spanned by evolving sources or literature values. Sources marked with a dagger ($\dagger$) have relativistic characteristic speeds and so may have to contend with the relativistic shock problem discussed in section~\ref{sec:shocks}. Sources marked with an asterisk ($*$) have strong magnetic fields and are subject to significant curvature or synchrotron losses which limit the maximum energy, so the Hillas energy is often a significant overestimate. A reference is given for each set of estimates, and the sub-section in which the source class is discussed is also labelled. The estimates presented here are inhomogenous and subject to large uncertainties and biases, and should not be taken as authoritative.
}
\label{hillas_table}
\end{table}

\section{Source Variability and UHECR Echoes ({\em When?})}
\label{sec:when}
`When' is perhaps an unusual interrogative to apply to the origin of UHECRs. Our motive for using it is to briefly describe our `echoes' model for UHECR production in dormant or declining sources, but time-dependence is an intrinsic part of any UHECR study, for a number of reasons. UHECR deflections in magnetic fields inevitably lead to a time delay with respect to any associated electromagnetic or neutrino signal \cite{lemoine_ultra--high-energy_1997,matthews_particle_2021}. Variability on a wide range of timescales is ubiquitous in accreting systems such as AGN \cite{uttley_brief_2004,maccagni_flickering_2020} and the fuelling of the AGN might be expected to cause flickering or stochastic activity \citep{gaspari_self-regulated_2016,yang_how_2016}. Star formation also varies over time due to various triggers and/or regulators \cite{beckmann_cosmic_2017,tacchella_stochastic_2020}, and some of the sources mentioned in the previous section are catastrophic events. With this in mind it seems reasonable to consider the time variability as an important factor for UHECR sources, as discussed before for transients like GRBs \cite{waxman_cosmological_1995,waxman_constraints_2009,globus_treasure_2022}.
\begin{figure}
\centering
\includegraphics[width=1.0\linewidth]{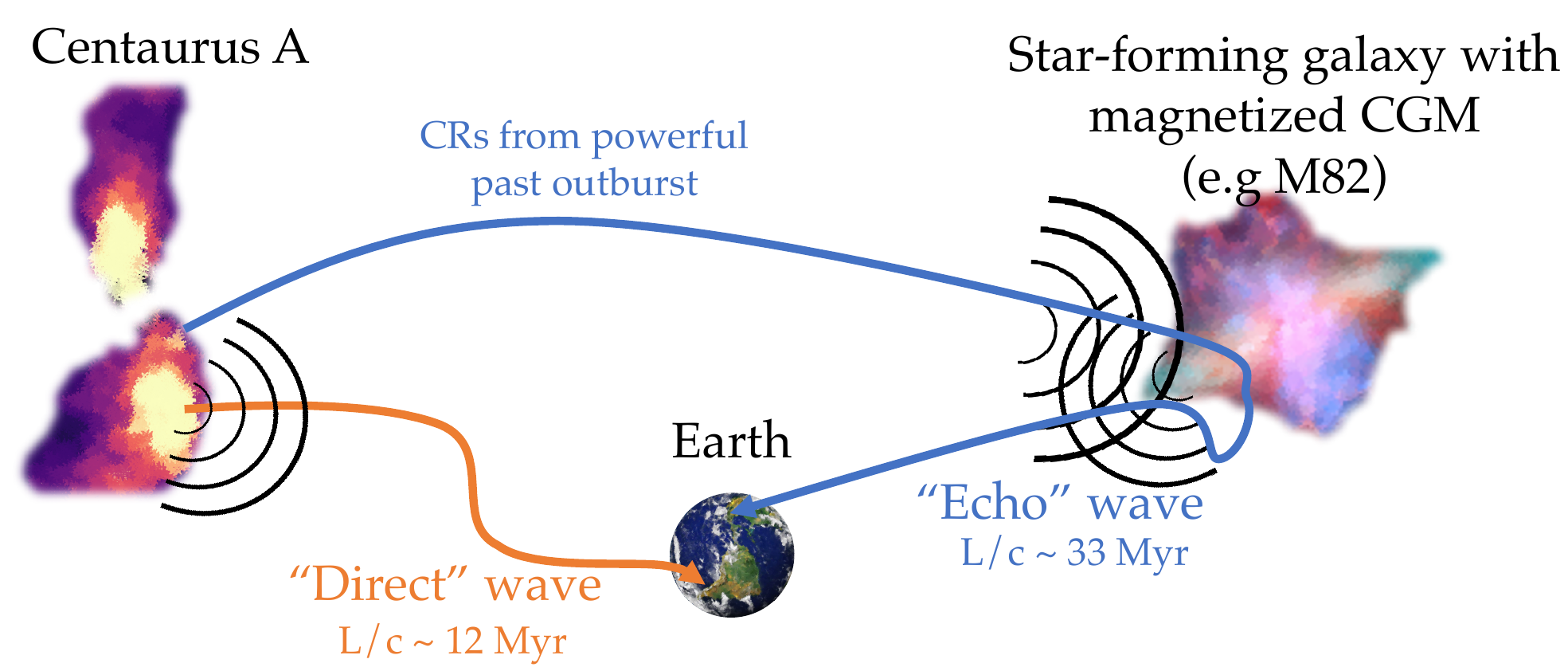}
\caption{
Schematic depicting the basic principle of the echoes model proposed by Ref. \cite{bell_echoes_2022} to explain intermediate scale anisotropies in PAO and TA data without requiring acceleration in starburst galaxies directly.
}
\label{fig:echo}
\end{figure}

Recently, Ref.~\cite{bell_echoes_2022} showed that a feasible model for intermediate UHECR anisotropies could be constructed, in which UHECRs are accelerated in a powerful outburst in a nearby source (Cen A), and the UHECRs then `echo' or scatter off nearby magnetic structures, in this case associated with the circumgalactic medium of star-forming galaxies that lie within a few Mpc in the CoG/Local Sheet. A schematic depicting this scenario is shown in Fig.~\ref{fig:echo}. The motivation here is that Cen A is unique in the CoG structure in having powerful AGN jet activity, and the model provides a way for arrival directions to be correlated with star-forming galaxies without requiring acceleration in the star-forming galaxies themselves. A workable model requires Cen A to have had a UHECR luminosity $20~{\rm Myr}$ ago that is $\sim200$ times that currently reaching Earth from the source, and also necessitates magnetic field strengths of $\sim 10-20~{\rm nG}$ out to a distance of $400-800~{\rm kpc}$ in the CGM. M82 could well be able to maintain a large magnetic field on this scale through advection or amplification of magnetic fields \cite{samui_efficient_2018,lopez-rodriguez_strength_2021,bell_echoes_2022}, in which case the UHECR echo could explain the TA hotspot. When star-forming galaxies such as NGC~253 and IC~342 are included, an anisotropy pattern can be produced that is similar to the all-sky anisotropy reported by PAO and TA \cite{biteau_covering_2019,di_matteo_full-sky_2020}.

The echoes model should be testable through the use of UHECR `composition clocks'. Rigidity-dependent propagation and species-dependent loss lengths (Fig.~\ref{fig:loss_lengths}) mean that, in principle, we should expect different compositions from the echo waves compared to the direct wave due to the different path lengths travelled (see Fig.~\ref{fig:echo}). In particular, species with short photodistintegration loss lengths could be under-represented in the echo wave, while high rigidity particles might be able to escape the source more quickly and be over-represented in the echo wave. We will present quantitative investigations of both of these effects in a future study (Taylor et al., in prep), which are exciting given the improved composition diagnostics of {\sl AugerPrime} \cite{augerprime}.

We note that many authors have discussed the impact of extragalactic magnetic field structures on the arrival directions of UHECRs. For example, Kotera \& Lemoine \cite{kotera_optical_2008} suggest  {\sl ``the possibility that the last scattering center encountered by a CR be mistaken with the source of this CR''}, while Kim et al. \cite{kim_filaments_2019} explore a similar effect encountered if UHECRs travel along magnetic filaments before scattering towards Earth. The deflection of UHECRs by magnetic fields is a generic difficulty associated with UHECR searches, which is exacerbated by how difficult it is to glean accurate knowledge of astrophysical magnetic field strengths and structures.

\section{Conclusions and Future Outlook}
\label{sec:conclusions}
The origins of UHECRs remain elusive, despite extensive efforts, and the study of their acceleration and propagation is a rich topic that has synergies with a whole host of subfields of astrophysics and particle physics. In this review, we have described how particles might be accelerated to super-EeV energies and discussed the basic energetic requirements for this to happen, informed in a large part by Hillas' 1984 work \cite{hillas_origin_1984}. We have touched on more detailed aspects of the plasma physics of particle acceleration, particularly relating to shock acceleration, that are critical to the particle acceleration process. We used these physical arguments to write a `checklist' for UHECR sources, which we applied to a range of astrophysical candidates such as AGN, GRBs, magnetised neutron stars and star-forming galaxies. Even in sources that dissipate energy at an astonishing rate, accelerating UHECRs still often requires particle acceleration physics to be stretched to its limits. 

As we have alluded to on multiple occasions in this review, we are presently in an exciting era for UHECR experiment and theory, as well as particle acceleration more generally. The UHECR spectrum is well-characterised, $X_{\rm max}$ distributions at ultrahigh energies are constraining the UHECR composition, and UHECR anisotropies are finally beginning to emerge above the noise at statistically significant levels. Nearly simultaneously, we have entered a `four-messenger' era of high-energy astrophysics through the observation, sometimes with electromagnetic counterparts, of $>$TeV neutrinos by IceCube \cite{ansoldi_blazar_2018} and gravitational waves by LIGO and VIRGO \cite{abbott_gravitational_2017}. In the future, the observational capabilities of the {\sl TA}x{\sl 4} \cite{tax4} and {\sl AugerPrime} \cite{augerprime} upgrades will dramatically improve our view of the UHECR sky, and the {\sl Cherenkov Telescope Array} will shortly offer unprecendented sensitivity to TeV gamma-rays \cite{cherenkov_telescope_array_consortium_science_2019}. In combination with a rapidly maturing theoretical landscape \cite{alves_batista_eucapt_2021}, these transformative instruments offer great prospects for fully understanding the `origin story' of UHECRs.

\section*{Acknowledgements}
\vspace{-0.5em}
We would like to thank Tony Bell, Lauren Rhodes, Frank Rieger, Claudio Galelli, Sergio Petrera and Alan Watson for helpful discussions. We are extremely grateful to J{\"o}rg Horandel and the organising committee for ECRS 2022 for an excellent conference. JM acknowledges funding from the Royal Society and, previously, from the Herchel Smith fund at Cambridge. 

\vspace{-0.5em}

\end{document}